\documentstyle[11pt]{article}
\begin{document}
\baselineskip 12pt
\parindent 10pt
\begin{center}
{\bf{Formation of the Ring-like Structure in the\\
SN 1987A Nebula due to the Magnetic Pressure\\
of the Toroidal Field}}\\
\end{center}
\begin{center}
 Haruichi Washimi$^1$, Shinpei Shibata$^2$ and Masao Mori$^3$\\
\vspace{8pt}
 $^1$Shonan Institute of Technology, Tsujido, Fujisawa, Kanagawa 251\par
 E-mail washimi@info.shonan-it.ac.jp\par
 $^2$Department of Physics, Yamagata University, Yamagata, Yamagata 990\par
 $^3$Department of Astronomy, School of Science, University of Tokyo,
 Bunkyo-ku, Tokyo 113\par
\vspace{6pt}
 (Received 1994 March 11; accepted 1995 October 14)\par
\end{center}
\baselineskip 20pt
\par
{\bf {Abstract}}\par
 Several weeks after the explosion of supernova (SN) 1987A, the UV flash of the SN illuminated a ring-like structure in the circumstellar material at about 0.65 ly from the SN. 
 \ The interaction between the stellar winds from the SN progenitor is considered to be the candidate for the formation of the circumstellar structure. 
 \ In the case that the stellar winds are spherically symmetric, the interaction should result in a shell-like structure. 
 \ However, in this paper we show that the magnetic field in the winds causes an anisotropy which leads to the formation of a ring-like structure. 
 \ When the fast wind of the blue supergiant phase of the progenitor sweeps up the surrounding slow wind of the red-supergiant phase, the magnetic field as well as the wind material are piled up in the interaction region. 
 \ Since the magnetic energy increases in proportion to the square of the amplitude, the magnetic field exhibits its effect prominently at the interaction region; due to the magnetic pressure force the material at lower latitudes is compressed into a ring-like structure. 
 \ It is suggested that this magnetic process can also explain the newly observed pair of rings of the SN 1987A nebula.\par
\vspace{6pt}
\newpage
{\bf{1. Introduction}}\par
 It is widely believed that the progenitor of SN 1987A was a red supergiant (RSG) which then evolved to a blue supergiant (BSG) just prior to the explosion. 
 \ In such a case, the circumstellar material surrounding the SN would comprise two types of stellar winds: 
 \ a dense and slow RSG generated wind far removed from the SN, and a dilute, fast BSG wind near to it. 
 \ When both winds are isotropic, as the BSG wind expands hydrodynamically into the RSG wind, a shell-like density enhancement is expected to form in the interaction region between the two winds. 
 \ Actual observations, however, do not support this simple picture. 
 \ Results from the European Southern Observatory (ESO, Wampler et al.~1990) and the Hubble Space Telescope (HST, Jakobson et al.~1991 and Panagia et al.~1991) reveal an unexpected ring-like structure, rather than a shell-like structure, at about 0.65 light-year (ly) from SN 1987A.\\

 In the early literature, ring-like structures were discussed in terms of planetary nebulae. 
 Moreover, such nebulae are also believed to generate a fast wind propagating into a slower wind.
 \ Several workers [e.g., Kwok (1982), Kahn and West (1985), Balick, Preston, and Icke(1987), Socker and Livio(1989)] have used hydrodynamic (HD) schemes to study the interaction region along with the assumption of a relatively high-density concentration along the equator of the slow wind. 
 \ They found that the interaction between the slow and fast winds enhances the density-anisotropy in the slow wind in such a way that  a ring-like structure is formed. 
 \ Based on this idea, Luo and McCray (1991) suggest that the ring-like structure of the SN 1987A nebula resulted from an anisotropy ratio of equatorial-to-polar density of 5 in the RSG wind (slow wind) before the interaction. 
\ Wang and Mazzali (1992) found that the outward propagating shell from the fast--slow boundary can encounter and substantially compress the anisotropy of a weak (ratio 1.2) slow-wind.\\

All of the above HD studies, however, seem to have neglected the pressure of the compressed gas at the interaction region, which would prevent the concentration of gas. 
 \ If this effect is taken into account, a much higher anisotropy ratio would be necessary for ring formation. 
 \ In fact, a recent full-HD computer simulation by Blondin and Lundqvist (1993) shows that a very high-density anisotropy of the order of 20 in the background RSG wind is required for ring formation. 
 \ In addition, a new image taken by the Hubble Space Telescope shows an additional pair of rings, which has a co-axis with the original ring (McCray, Lin 1994). 
 \ This new image seems to be inaccessible in terms of the HD point of view. \\

 In this paper we propose an MHD model in which the magnetic pressure effect can account for the formation of the ring-like structure. 
 \ The main purpose is to explain the original ring, but as will be shown below, a pair of rings is also reproduced in our MHD simulation, though the geometrical scale should be adjusted by a further fine tuning of the model. \\

 Based on an evolutionary scenario of the progenitor's magnetic field and rotation, the RSG wind is likely to posess such a magnetic field strength and rotation rate that the magnetic pressure causes a density anisotropy. 
 \ It is shown for a magnetic stellar wind from a dipole source field (Washimi, Shibata 1993) that the circumstellar poloidal magnetic field ($B_p$, the field on the meridional plane) configuration becomes a split monopole configuration; the field is oriented to the radial direction, but its polarity changes in the equatorial region, and, thus, a magnetic neutral sheet is formed along the equator. 
 \ The toroidal magnetic field ($B_{\phi}$, the field of the toroidal component around the stellar rotation axis, or in other words, the field vertical to the meridional plane) is proportional to  $\Omega \cdot B_p \cdot R \cdot sin\ \theta $, where $\Omega$ is the stellar rotation speed, $R$ the distance from the star, and $\theta$ the colatitude angle. 
  \ Since $B_p$ and $B_{\phi}$ decrease as $R^{-2}$ and $R^{-1}$, respectively, $B_{\phi}$ is much stronger than $B_p$ in circumstellar space far from the stellar surface. 
 \ Thus, a thin magnetic neutral sheet of the toroidal field is formed around the equator where the density is enhanced to maintain the total pressure balance, as shown by Washimi and Shibata. 
 \ During the dynamical stage of the sweeping up of the RSG wind by the BSG wind, the toroidal field is piled up in the interaction region.
 \ Since the energy increases as the square of the intensity, this piling up leads to a drastic increase in the magnetic energy. 
 \ Due to all of the above reasons, the magnetic pressure force of the toroidal field compresses the gas (plasma) at the equatorial interaction region and maintains a localized high-density distribution over a long-time period by pressure balancing the ring of the enhanced gas. \\

{\bf{2. Basic Model and Simulation Method}}\par
 We perform an MHD computer analysis in order to confirm this physical process. Both the RSG and BSG winds are assumed to be obeyed by the ideal MHD system of equations, i.e.,\\
\par
\leftskip 0.8cm
\begin{tabular}{p{8.2cm}c}
${\partial\rho/\partial t}+{\bf \nabla\cdot({\rho}{\bf v})}=0,$ &(1) \\
${\partial{\bf v}/\partial t}+({\bf v}\cdot {\bf\nabla}){\bf v}=-(1/\rho){\bf \nabla} P +(1/\rho){\bf J}\times{\bf B},$ &(2)\\
${\partial P/\partial t}+({\bf v}\cdot{\bf \nabla})P = -\gamma P{\bf \nabla}\cdot{\bf v},$ &(3)\\
${\partial{\bf B}/\partial t} = {\bf\nabla}\times({\bf v}\times{\bf B}), $ &(4)\\
and\\
$\mu_0 {\bf J} ={\bf \nabla}\times{\bf B},$ &(5)\\
\end{tabular}
\par
\leftskip 0cm
\vspace*{4mm}
 where $\mu_0$ is the permeability and the polytropic index $\gamma$ is 5/3.
 \ Here, the cylindrical coordinates ($\it{r}$, $\phi$, $\it{z}$) with the $\it{z}$-axis aligned along the rotation axis of the progenitor, are used and axial symmetry is assumed. The flow velocity, {\bf {v}}, the magnetic field, {\bf {B}}, and the current, {\bf {J}}, in equations (1)-(5) are three-dimensional vectors.\\

  Once a magnetic field is introduced, we may have a wide variety of parameters for the wind structure, the magnetic field configurations, and the latitudinal dependence of the wind velocity and density.  
 \ In the present analysis, we attempt to show the ring formation using the very simple, but reasonable, parameters, as follows. 
 \ The mass-loss rate, speed, and the temperature are $10^{-5}\ M_\odot yr^{-1}$, $V^{RSG}=10\ km\ s^{-1}$ and  $5.0\times 10^3\ K$ (at $0.1$ ly), respectively for the RSG wind, and are $3.0\times 10^{-6}M_\odot\ yr^{-1}$, $V^{BSG}=550\ km\ s^{-1}$ and $1.3\times 10^6\ K$ (at $0.1$ ly), respectively, for the BSG wind. 
 \ Except for the temperature of the BSG wind, the other parameters are similar to those of Shigeyama and Nomoto (1990) and Luo and McCray (1991). 
 \ The relatively high temperature of the BSG wind derives from a technical convenience to sustain the Rankine-Hugoniot shock condition at the reverse shock in our computation. 
 \ Since the ram pressure is much higher than the thermal pressure in the case of BSG wind, this choice does not lead to any unphysical results. 
 \ The radius and the rotation speed of the progenitor are chosen to be  $R_0^{RSG}=3.1 \times 10^{11}\ m$ and $\Omega ^{RSG}=3.5 \times 10^{-9}\ rad\ s^{-1}$, for the RSG star; whereas $R_0^{BSG}=4.9 \times 10^{10}\ m$ and $\Omega ^{BSG}=3.5 \times 10^{-8}\ rad\ s^{-1}$ for the BSG star. 
 
 Indeed, a realistic estimate of the angular velocity as well as the magnetic field strength of the progenitor is very important for ring formation.
 \ Therefore, we need a detailed consideration of the stellar evolution for choosing these parameters. 
 \ In general, main-sequence stars bluer than the spectral type F5 are known as rapid rotators (typically $V_{eq} \approx$ 100 $km\ s^{-1}$), while redder stars are slow rotators (Tassoul 1978). 
 \ This change in rotation along with spectral types is also observed for sub-giants and giants, where the change is seen around G0 type (for supergiants, it is obscured by the convection motion). 
 \ This fact suggests that the on-set of surface convection triggers the dynamo, and then a strong magnetic breaking of the stellar rotation. 
 \ We can thus infer a dynamo-onset-line on the HR diagram.\\

 As a main-sequence star, the progenitor would rotate at typically $R_0\Omega \sim  100\ km\ s^{-1}$ ($\Omega \sim 2.5\times 10^{-5}\ rad\ s^{-1}$, and $R_0 \sim 4\times 10^9\ m$), where $R_0$ is the radius of the progenitor, which changes with stellar evolution. 
 \ As the progenitor evolved towards RSG, it span down due to an increase in the moment of inertia. 
 \ Taking the moment of inertia given by Eriguchi et al. (1992) and assuming the conservation of angular momentum, we have $R_0\Omega \sim 3\ km\ s^{-1}$ ($\Omega \sim 10^{-8}\ rad\ s^{-1}$), which is consistent with the observed rotation velocity of typical supergiants (Gray 1991). 
 \ According to the suggested evolution (Saio et al. 1988), the progenitor passed the dynamo-onset-line just at its RSG phase, and would suffer magnetic breaking. 
 \ If $\bigtriangleup M$ is the lost mass after the development of the dynamo, the corresponding loss of angular momentum is given by $\bigtriangleup L = \bigtriangleup M \Omega R_A^2$, where $R_A$ is the averaged radius of the Alfven surface. 
 \ The angular momentum of the star is denoted by $L = fMR_0^2\Omega$, where $f \sim 4.6 \times 10^{-2}$ (Eriguchi et al. 1992). 
 \ Therefore, a prominent spin-down takes place when $\bigtriangleup L/L = (\bigtriangleup M/0.1M_{\odot})(R_A/2.7R_0)^2 \sim 1$, i.e., when the magnetic field is amplified so that $R_A$ increases up to $\sim 2.7 R_0$. 
 \ At this stage, the corotating velocity $R_A\Omega$ becomes $\sim$ 8\ $km\ s^{-1}$, which is comparable with the expected velocity of the RSG wind. 
 \ Therefore the centrifugal driving force is as important as other driving forces, such as due to the Alfven wave. 
 \ An immediate corollary of this fact is that the magnetic pressure $B_{\phi}^2/2\mu_0$ beyond the Alfven surface is comparable with the dynamic pressure.\\
 
 Thus, it is very likely that the magnetic pressure produces a density anisotropy in the RSG wind. 
 \ For a numerical simulation, we regard a suggestion based on the synchrotron radio emmission of SNR 1987A, and that the ring material was emitted at the later phase of RSG; we thus chose a set $B_p = 12 (R_0/R)^2$ Gauss and $\Omega = 3.5\times 10^{-9}\ rad\ s^{-1}$, for which the ratio of the magnetic energy to the flow energy is $\sim$ 0.9. 
\ By using this set of $B_p$ and $\Omega$, we obtain a field strength on the order of $10^{-4}\ $Gauss in the shocked gas, which is suggested by the synchrotron emission (Staveley-Smith et al. 1992). 
 \ Another crude estimate of the field strength on the RSG star surface may be provided by equipartition. 
 \ According to the projenitor model by Saio et al. (1988), three of the radiation-energy densities, the convectional one and the thermal one of the gas, are comparable to each other. 
 \ If the magnetic energy is equated with these densities ($B^2 / 8 \pi \approx \epsilon_{\rm conv} \approx a T^4 \approx n k_B T$) we have $\approx 10\ $Gauss.
 \ However, any other set of the field strength and the angular velocity with the similar energy ratio could be taken for the ring formation. 
 \ The important point is that the condition for the density anisotropy is equivalent to the condition for the magneto-centrifugal driven wind, which is very likely to have occurred in the RSG phase. \\

 The poloidal magnetic field of both the RSG and BSG winds is assumed to be a split monopole followed by the previous discussion; 
 \ the field is radially oriented and the colatitudinal dependence of the field is assumed to be $B_p=b_p/R^2$ for $0^{\circ} \leq \theta \leq \theta_0$, where $b_p$ is a constant, with a decrease to zero on the equator, so that $B_p=b_p\cdot (\theta-90^{\circ})/(\theta_0-90^{\circ})/R^2$ for $\theta_0 < \theta \leq 90^{\circ}$, and changes its sign in the southern-hemisphere. 
 \ We assume $\theta_0=85^{\circ}$ in this simulation. 
 \ The BSG poloidal field is prescribed by the magnetic flux conservation: $b_{p0}^{BSG}=(R_0^{BSG}/R_0^{RSG})^2 \cdot b_{p0}^{RSG}$. 
 \ The associated toroidal magnetic field is equal to $(\Omega^{RSG}\cdot R/V^{RSG})\cdot B_p^{RSG}\cdot sin\ \theta$ in the RSG wind, and to $(\Omega^{BSG}\cdot R/V^{BSG})\cdot B_p^{BSG}\cdot sin\ \theta$ in the BSG wind.
 \ It may be notable here that the magnetic field is virtually toroidal, as can be seen from the above. 
 \ The poloidal field is truly minor and follows the meridional flow, which is, however, affected by the toroidal field distribution.

 Under these conditions, the ram pressure $\rho v^2$, the thermal pressure $P$, and the magnetic pressure $B_\phi ^2/2\mu_0$ at $R=0.1$ ly are $5.6\times 10^{-10}$, $2.3\times 10^{-10}$, and $7.1\times 10^{-10}$ respectively, expressed in units of $N\cdot m^{-2}$ for the RSG wind, and $9.3\times 10^{-9}$, $3.3\times 10^{-10}$, and $2.3\times 10^{-11}$, respectively for the BSG wind.\\

 There are $960 \times 960$ grid points for physical domain $1$ ly $\times 1$ ly. 
 \ For the initial condition, the isotropic RSG wind, the split-monopole poloidal and the associated toroidal magnetic fields are set in a whole simulation box, except on the inner boundary, which is located at $R_{IB} = 0.1$ ly from the SN. 
 \ The isotropic BSG wind and the BSG poloidal and toroidal magnetic fields are prescribed on the inner boundary. 
 \ On this boundary, the BSG poloidal field is the same as the RSG one, while the BSG toroidal field is prescribed by the relation, $(\Omega^{BSG}\cdot R_{IB}/V^{BSG}) \cdot B_p^{BSG} \cdot sin\ \theta$. 
 \ Although this toroidal field is smaller than the RSG one, some strong toroidal field is expected to be accumulated at the contact surface between the RSG and BSG winds. 
 \ The time evolution of the interaction between the two winds is computed step by step using the two-step Lax-Wendroff scheme.\\

{\bf{3. Computational Results and Conclusions}}\par
 Figures 1 and 2 show the global patterns of the density and magnetic field intensity, respectively, when the forward shock propagates up to $\approx$ 0.65 ly. 
 \ A finite-density enhancement along the equator due to the equatoward magnetic pressure force is found in the RSG wind in this figure (and also in figure 3). 
 \ This enhancement is a natural consequence due to a magnetic-pressure effect, which contrasts to some enhancements under an ad hoc assumption in HD schemes by Luo and McCray (1991) and Wang and Mazzali (1992). 
 \ It is worth noting that the equatorial enhancement is not found in the BSG wind, where the toroidal magnetic field is relatively weak. 
 \ A dramatic magnetic-pressure effect is found at around the contact surface where the BSG wind sweeps up the RSG wind. 
 \ That the density increases at the shell-like structured region of the interaction between the contact surface and the forward shock over all latitudes has been confirmed. 
 \ The special sharp enhancement, which corresponds to the observed ring-like structure, is found on the equator. 
 \ Its peak density is $3.2\times 10^3\ cm^{-3}$.
 
 Here, the role of the magnetic pressure is emphasised; if this pressure is absent the accumulated RSG wind matter is not confined in a ring-like strucuture. 
 \ Figure 2 shows that due to the sweeping-up effect of the BSG wind the magnetic field is also accumulated at around the contact surface. 
 \ It is remarkable that the magnetic intensity decreases drammatically to zero at the equatorial edge of the shell, where the sharp density-enhancement is found in figure 1. 
 \ This means that the surrounding magnetic pressure pushes and confines the RSG material to the equator from both the northern and southern sides, and thus the ring-like structure is formed. 
 \ Between the reverse shock located at $\sim$ 0.27 ly and the contact surface near to the forward shock we find a hill-like widespread magnetic field accumulation of the shocked BSG toroidal field. 
 \ As is shown below, the magnetic pressure here also works on the BSG wind. \\

 In this paper our attention has been focused on the magnetic effect around the equatorial plane. 
 \ Here, we note that, in addition to the equatoward force in low latitudes, there is also a poleward force of the magnetic pressure in the middle and high latitudes which results in a density enhancement around the polar axis (Washimi, Shibata 1993). 
 \ Because the latitudinal dependence of the toroidal field is $B_\phi \propto sin\ \theta$ for $0^{\circ} \leq \theta \leq \theta_0$ in the initially prescribed RSG wind, the magnetic pressure also pushes the RSG wind poleward in the middle and high latitudes. 
 \ In fact, due to this poleward pressure force along the shell-like enhancement, the density increases with latitude in the middle latitudes, 
 \ but it decreases again close to the polar region, and thus a local density maximum of $\sim 1.6 \times 10^3\ cm^{-3}$ is found at $\theta \sim 15^{\circ}$. 
 \ This is probably due to a dynamical effect of the BSG wind (see figure 3), which prevents a density accumulation just on the pole. \\

{\bf{4. Discussion}}\par
 In this MHD simulation analysis we have obtained an enhanced density of $\approx 3.2 \times 10^3\ cm^{-3}$ at the ring. 
 \ This peak value will probably be revised to a greater one by a more sophisitcated analysis, and will be consistent with the observed density $(2$--$3)\times 10^4\ cm^{-3}$ (Lundqvist, Fransson 1991).
 \ A suppression of the peak density value could probably be caused by the numerical diffusion effect, although we have attempted to reduce this effect by using a large mesh number. 
 \ In addition, if our calculation could be performed from very near to the progenitor, where the poloidal field would be a dipole-like, rather than a split monopole-like configuration, the equatorward concentration of the material in the RSG wind would occur in a much greater colatitudinal width (e.g., 45$^{\circ}$--135$^{\circ}$), because the amplitude of the associated toroidal field ($\propto \sin \theta \cdot \cos \theta$) is maximum at $\theta=45^{\circ}$ and $135^{\circ}$. 
 \ For that case, the equatorial enhancement would be greater than our computational value, for which the split-monopole of the angle 85$^{\circ}$--95$^{\circ}$ is conventionally used. In addition, if some cooling effect could be taken into account in the simulation, the density enhancement could be larger. \par

 We note that the idea of a magnetic field effect is consistent with the recent radio observation of a supernova remnant, SNR 1987A, detected by Staveley-Smith et al. (1992) at about 1200 days after the explosion. 
\ This radio emission is explained by the collision of the supernova blast wave with the shocked blue wind (reverse shock at about 0.27 ly where some magnetic field increase and a small density change are found in figures 2 and 1, respectively). 
 \ This position corresponds to the averaged expansion speed of the supernova ejecta $\sim  0.08$ ly which is consistent with the estimation by Shigeyama and Nomoto (1990). 
 \ The estimated magnetic-energy density by the minimum-energy argument is $\sim 4 \times 10^{-8} f^{-4/7}\ N\ m^{-2}$, where $f$ is the fractional volume of the radiating acceleration region, suggesting a magnetic field of a few milli-Gauss or more (Chevalier 1992; Ball, Kirk 1992). 
 \ This field intensity is consistent with an intensity of $\approx 2\cdot 10^{-4}$  Gauss obtained between the reverse shock and the contact surface shown in figure 2, if one takes into account a further enhancement of the field due to the sweeping-up process by the supernova blast wave. 
 \ When the SN ejector collides with the ring at the end of this century or at the beginning of the next one, we can also expect more intense radio emission at rather middle and high latitudes where the magnetic intensity is greater, rather than at the equator where the ring-like structure is located.\\

 In addition, we note that the expanding speed of the ring in our present simulation is 32$\ km\ s^{-1}$ which is higher than the estimated speed of 10.3$\ km\ s^{-1}$ (Crotts, Heathcote 1991). 
 \ The resulting speed is a natural consequence of our parameters; we have used some conventional speeds of the RSG and BSG winds, 10 and 550$\ km\ s^{-1}$, respectively. 
 \ We guess that these values should be lower to explain the ring speed.\\

Chevalier and Luo (1994) studied the magnetic effect in the formation of the ring. 
 \ Their idea is that the magnetic pressure and tension deform the shell to give an hourglass shape and the ring. 
\ However, their model is too simplified.
 \ Since both the latitude-dependence of the toroidal field strength, which has been shown to be essential for ring formation, and the inertial effect are not taken into account, their dynamical process for the formation of the ring-like structure is much different from ours. \\ 

 It is interesting to note that the local-density maximum around the pole axis in our simulation constitutes additional rings in both the northern and southern hemispheres, which can corespond to the newly observed pair of rings. 
 \ The opening angle $\sim 15^{\circ}$ and the diameter $\sim 0.15\ ly$ are  different from the observed values ($\sim 45^{\circ}$, $\sim 1.4\ ly$);\ however, these quantitative discrepancies could be overcome in future analyses if some conditions, such as high-speed wind flow in high latitudes, as observed in the solar wind, are additionally taken into account. \\

\vspace{10pt} 
 We wish to thank Dr. Saio for valuable discussions concerning stellar evolution, and to Drs. P.K. Manoharan and A. Sterling for a critical reading of the manuscript. This work was supported by a Grant-in-Aid for Scientific Research on Priority Areas from the Ministry of Education, Science, Sports and Culture of Japan and by the Collaboration Study Program of the STE Laboratory. The computations were performed on the parallel supercomputer, ADENART-256.\\
\par
{\bf {References}}\par
\leftskip 0cm
\leftskip 4ex
\parindent -4ex
\ \par
Balick B., Preston H.L., Icke V. 1987, AJ 94, 1641
\parindent -4ex
\ \par
Ball L., Kirk J.G. 1992, ApJL 396, L39
\parindent -4ex
\ \par
Blondin J. M., Lundqvist P. 1993, ApJ 405, 337
\parindent -4ex
\ \par
Chevalier R.A. 1992, Nature 355, 617
\parindent -4ex
\ \par
Chevalier R.A., Luo, D. 1994, ApJ 421, 225
\parindent -4ex
\ \par
Crotts A.P.S., Heathcote S.R. 1991, Nature 350, 683
\parindent -4ex
\ \par
Eriguchi Y., Yamaoka H., Nomoto K., Hashimoto M. 1992, ApJ 492, 243
\parindent -4ex
\ \par
Gray D.F 1991, in Angular momentum evolution of young stars ed S.R. Catalano, NATO ASI Ser. C, 340, p183
\parindent -4ex
\ \par
Jakobson P., Albrecht R., Barbieri C., Bkades J.C., Boksenberg A., Crane P., Deharveng J.M. et al. 1991, ApJL 369, L63
\parindent -4ex
\ \par
Kahn F.D., West K.A. 1985, MNRAS 212, 837
\parindent -4ex
\ \par
Kwok S. 1982, ApJ 258, 280.
\parindent -4ex
\ \par
Lundqvist P., Fransson C. 1991 ApJ 380, 575
\parindent -4ex
\ \par
Luo D., McCray R. 1991, ApJ 379, 659
\parindent -4ex
\ \par
McCray R., Lin D.N.C. 1994, Nature 369, 378
\parindent -4ex
\ \par
Panagia N., Gilmozzi R., Macchetto F., Adorf H.-M., Kishner R.P. 1991, ApJ 380, L23
\parindent -4ex
\ \par
Saio H., Kato, M., Nomoto K. 1988, ApJ 331, 466
\parindent -4ex
\ \par
Shigeyama T.,  Nomoto K. 1990, ApJ 360, 242
\parindent -4ex
\ \par
Socker N., Livio M. 1989, ApJ 339, 268
\parindent -4ex
\ \par
Staveley-Smith L., Mancheste R.N., Kesteven M.J., Campbell-Wilson D., Crawford A.J., Turtle A.J., Reynolds J.E., Tziomis A.K. et al. 1992, Nature 355, 147
\parindent -4ex
\ \par
Tassoul J.L. 1978, Theory of rotation stars (Princeton University Press, Princeton), p28
\parindent -4ex
\ \par
Wang L.,  Mazzali P.A. 1992, Nature 355, 58
\parindent -4ex
\ \par
Wampler E.J., Wang L., Baade D., Banse K., D'Odorico S., Gouiffes C., Tarenghi, M. 1990, ApJL 362, L13.
\parindent -4ex
\ \par
Washimi H.,  Shibata S. 1993, MNRAS 262, 936
\parindent -4ex
\ \par
\newpage
{\bf {Figure Captions}}
\parindent -4ex
\ \par
Fig. 1. Global density pattern for the region outside $R=0.1$ ly (inner boundary) when the forward shock reaches  $\approx 0.65$ ly. \ The shell-like structure of the enhanced density is found, as well as the sharp peak on the equator corresponds to the ring-like structure. 
 \ Its peak value is $\approx$ 3,200$\ cm^{-3}$, while the peak value at $\theta \approx 15^{\circ}$ is $\approx$ 1,600$\ cm^{-3}$. The positions of the reverse shock and the inner boundary are indicated by `RS' and `IB', respectively.
\parindent -4ex
\ \par
Fig. 2. Global toroidal magnetic field intensity pattern for the region outside $R=0.1$ ly (inner boundary) when the forward shock reaches $\approx 0.65$ ly. 
 \ Note the great field accumulation in the shocked RSG wind. The sharp decrease in the field accumulation appears at the equator, just where the sharp density peak in figure 1 is located. 
 \ A hill-like accumulation of the magnetic field is also found between the contact surface and the reverse shock. 
 \ The positions of the reverse shock and the inner boundary are indicated by `RS' and `IB', respectively.
 \ The plot does not show the exact equatorial plane where the toroidal field vanishes, but shows down to the grids just above the equator.
\parindent -4ex
\ \par
Fig. 3. Equi-density pattern (solid curves) and wind velocity (arrows).
 \ The positions of the forward and reverse shocks and the inner boundary are indicated by `FS', `RS' and `IB', respectively.
 \ The highest contour level is $10^{3.5}$cm$^{-3}$, and the contour levels go down with the step 0.1 in logarithmic scale.
\parindent -4ex
\ \par

\newpage
\begin{figure}[p]
\vspace*{20cm}
\includegraphics{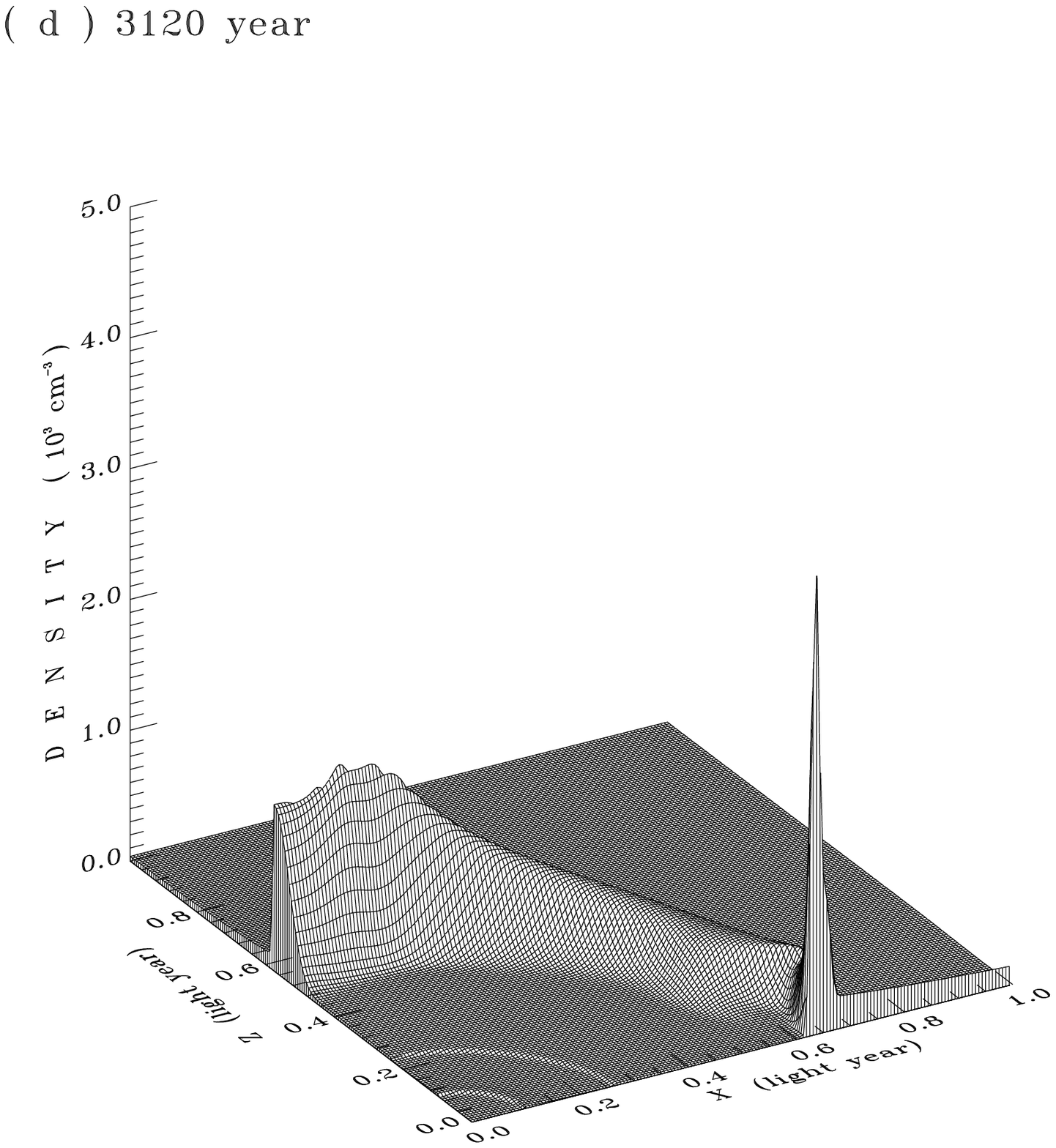}
\end{figure}
\newpage
\begin{figure}[p]
\vspace*{20cm}
\includegraphics{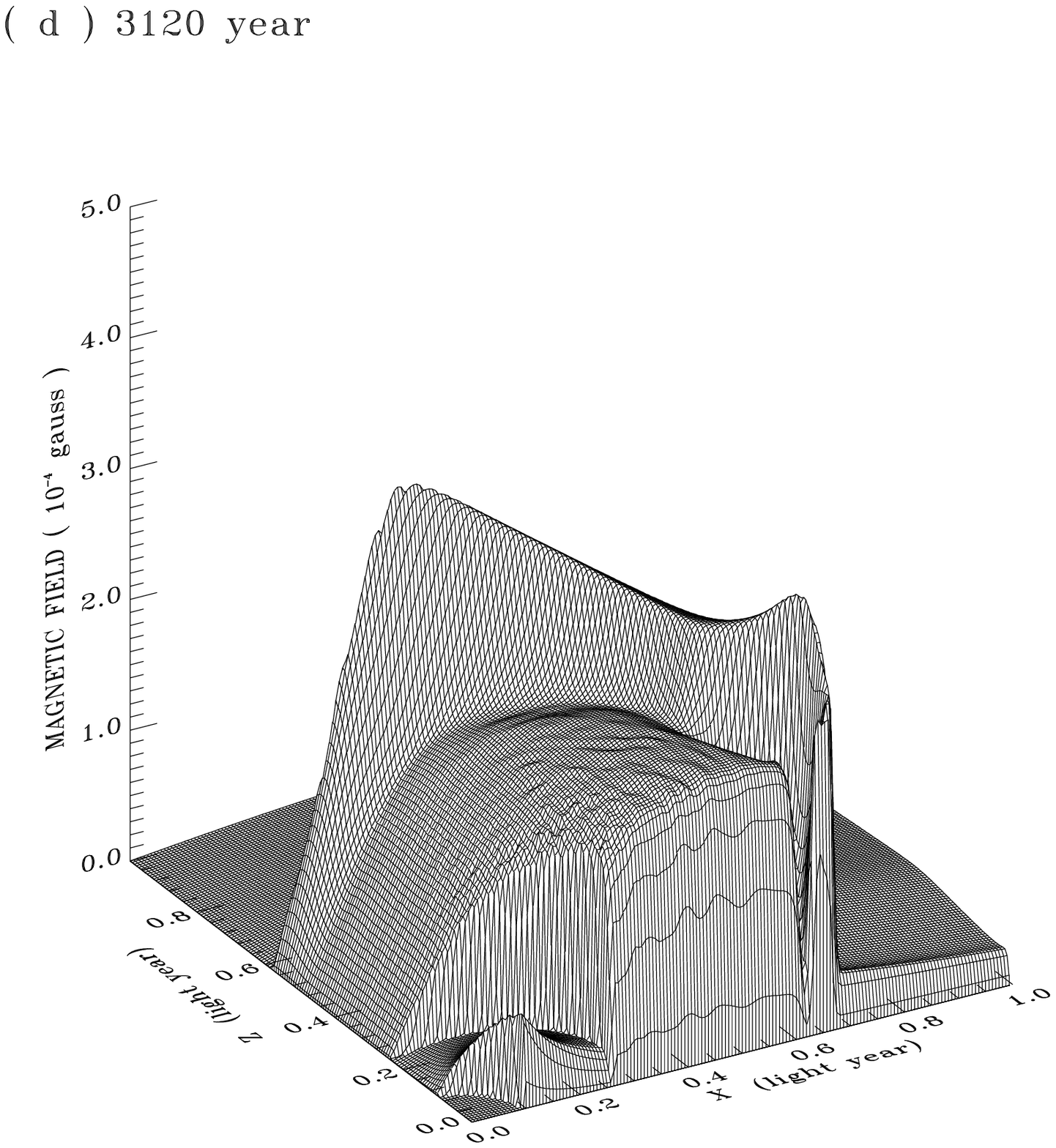}
\end{figure}
\newpage
\begin{figure}[p]
\vspace*{20cm}
\includegraphics{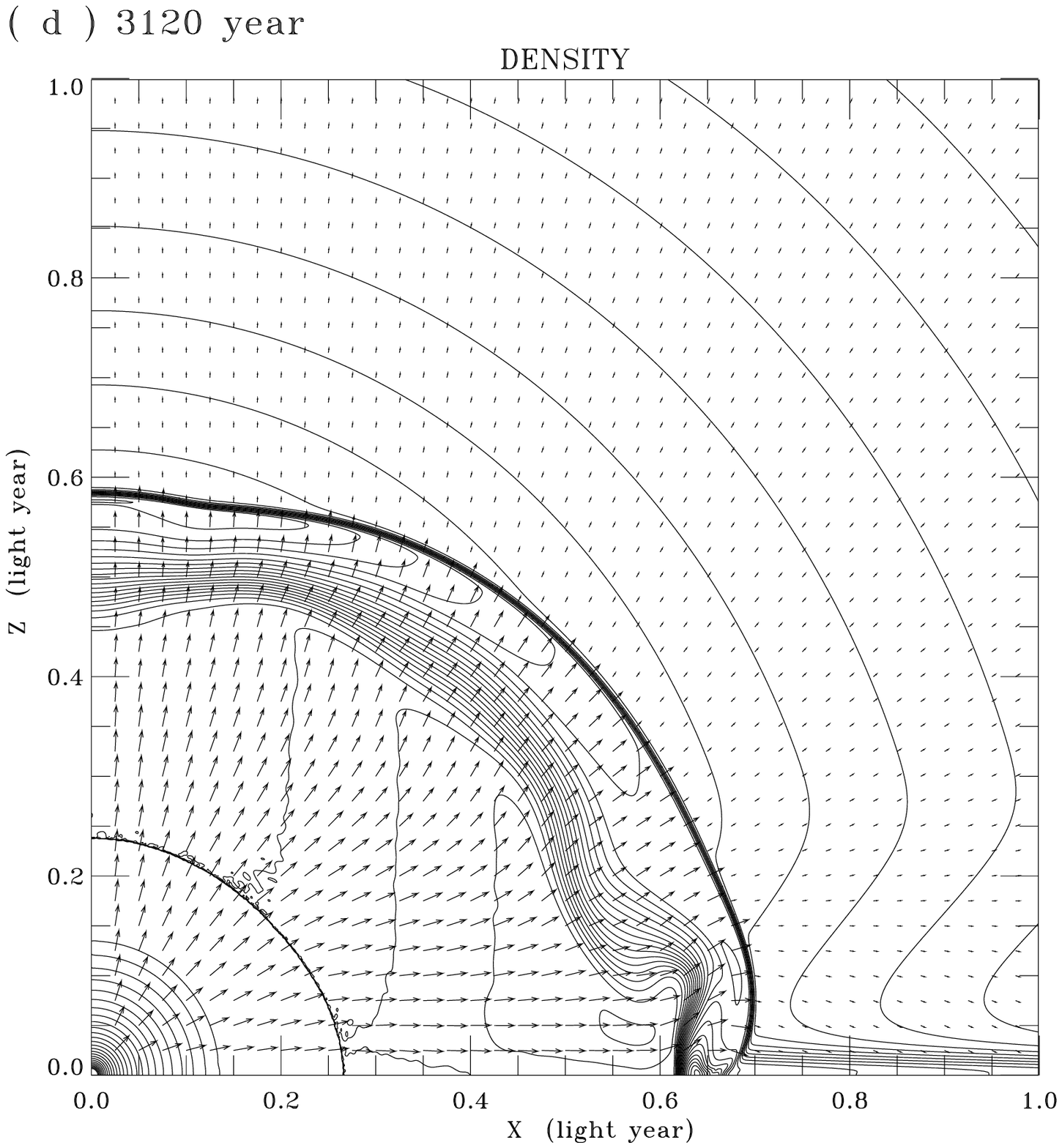}
\end{figure}

\end{document}